# Efficient synthesis of Vitamin D₃ in a 3D ultraviolet photochemical microreactor fabricated using an ultrafast laser

Aodong Zhang[1,2,3], Jian Xu[1,2,3,*], Lingling Xia[1], Ming Hu[1,*], Yunpeng Song[2,3], Miao Wu[3], and Ya Cheng[1,2,3,4,*]

[1] Engineering Research Center for Nanophotonics and Advanced Instrument, School of Physics and Electronic Science, East China Normal University, Shanghai 200241, China;
[2] State Key Laboratory of Precision Spectroscopy, School of Physics and Electronic Science, East China Normal University, Shanghai 200241, China;
[3] XXL—The Extreme Optoelectromechanics Laboratory, School of Physics and Electronic Science, East China Normal University, Shanghai 200241, China;
[4] State Key Laboratory of High Field Laser Physics, Shanghai Institute of Optics and Fine Mechanics, Chinese Academy of Sciences, Shanghai 201800, China

**Abstract**

Large-scale, high-precision, and high-transparency microchannels hold great potential for developing high-performance continuous-flow photochemical reactions. We demonstrate ultrafast laser-enabled fabrication of 3D microchannel reactors in ultraviolet (UV) grade fused silica which exhibit high transparency under the illumination of UV light sources of wavelengths well below 300 nm with excellent mixing efficiency. With the fabricated glass microchannel reactors, we demonstrate continuous-flow UV photochemical synthesis of vitamin D₃ with low power consumption of the UV light sources.

**Keywords:** Ultrafast laser microfabrication; fused silica microchannels, UV photochemical reactors, continuous-flow synthesis, vitamin D₃

*Correspondence: jxu@phy.ecnu.edu.cn; mhu@phy.ecnu.edu.cn; ya.cheng@siom.ac.cn

# Introduction

Vitamin $D_3$ (VD3) has numerous important applications in biomedicine and clinical therapy, consequently creating a substantial market demand for its large-scale and high-purity production[1-5]. In nature, VD3 can be photochemically generated in the bodies of animals and humans from 7-dehydrocholesterol (7-DHC) initiated by proper ultraviolet (UV) light irradiation (particularly within the spectral range from ~275 nm to ~300 nm). Currently, the industrial synthesis of VD3 predominantly relies on photochemical reactions, wherein 7-DHC is irradiated by UV light to yield pre-vitamin $D_3$, subsequently thermally isomerizing into VD3. However, traditional photochemical synthesis conducted in batch photoreactors suffers from low photon utilization efficiencies, thereby limiting the 7-DHC conversions. The yield of VD3 is typically less than 20%, giving rise to low production efficiency, high production cost, and unwanted byproducts and isomers caused by long-term irradiations. The emergence of microfluidic technology and the integration of continuous-flow chemical reactions into photochemical synthesis have ushered in new prospects for fine chemicals and pharmaceutical synthesis[6-13]. Although continuous-flow chemical synthesis has proven its prominent advantages for drug and drug intermediate production, current research and application of continuous-flow photochemical microreactions primarily utilizes visible light (400~700 nm) and medium-to-long-wavelength UV light (300~400 nm) as irradiation sources[9, 12,13]. To date, UV-irradiation continuous-flow chemical synthesis of VD3 faces several critical limitations on microreactor materials and UV light sources. Regarding the microreactor materials, conventional transparent UV polymer-based microreactors, such as fluorinated ethylene propylene (FEP) tubes[14,15], inevitably degrade under the long-term operation of UV light irradiation and high temperatures. Moreover, current commercially available glass microchannel reactors (typically made of borosilicate glass)[11, 12, 16, 17] exhibit poor UV light transmission rates in the short-wavelength UV spectrum, hindering high-efficiency photon utilization for UV photochemical synthesis. Fused silica glass, renowned for its superior UV transparency ranging from 200 nm to 400 nm, represents an ideal microreactor material for such reactions. However, the fabrication of microreaction chips in fused silica is challenging due to the difficulties in microfabrication and high-performance bonding of fused silica glass[18].

The existing fused silica glass microchannel reactors primarily comprise glass capillaries of uniform circular cross-sections, which is not ideal for achieving efficient fluid manipulation and high photon irradiation efficiency. Furthermore, most UV continuous-flow photochemical microreactors employ medium- and high-pressure mercury lamps as UV light sources which consume power significantly[2, 3, 5, 19]. In some cases, polymer tubes or fused silica capillaries are coiled around these lamps, inevitably generating excess heat during the long-term operation of the reactors due to the high photon energies associated with UV lights. This necessitates stringent demands on the temperature control system throughout the reactor. To overcome the difficulties, the development of a new UV-light-enabled continuous-flow photochemical fused silica microchannel reactor system is highly desirable.

Herein, we propose a hybrid laser microfabrication approach for the manufacture of UV photochemical fused silica microchannel reactors, which is based mainly on the ultrafast laser microfabrication of fused silica assisted by carbon dioxide ($CO_2$) laser processing for sealing the extra-access holes left behind the ultrafast laser microfabrication. In this approach, three-dimensional (3D) embedded micropatterns in fused silica, including the patterns of 3D microchannels and a string of extra-access holes between the channels and glass surfaces, are first created by ultrafast laser direct writing[20-22]. The distribution of extra-access holes is carefully optimized to improve the subsequent etching homogeneity of the laser-modified structures[22-24]. After chemical etching, the holes are sealed by defocused $CO_2$ laser irradiation to form closed microchannel structures with a few inlets and outlets for introducing chemical reagents into the microchannels and extracting the products out of the microreactor. With the manufactured microchannel reactors, we demonstrate continuous-flow UV photochemical synthesis of vitamin $D_3$ in a one-step and low-power-consumption manner.

## Results and discussion

### Manufacture of a 3D large-scale fused silica microchannel reactor

The manufacture of a 3D large-scale fused silica microchannel reactor is based on the hybrid laser microfabrication, which combines the merits of ultrafast laser-assisted etching, extra-access holes for enhancing the etching, and $CO_2$ laser-induced in-situ melting of glass surface for hole sealing. As illustrated in Figure 1a, the manufacturing flow for a 3D large-scale fused silica microchannel structure consists of three main procedures: (i) ultrafast laser direct writing of 3D microchannels comprised of 3D micro-mixing units and a string of vertical extra-access holes distributed evenly along the channel; (ii) selective chemical etching for removing the laser-modified micropatterns to obtain a hollow microchannel with a string of through holes reaching the glass surface; (iii) $CO_2$ laser irradiation for sealing the holes to form a closed large-scale microchannel structure with a few necessary inlets and outlets. First, an ultrafast laser beam was focused through an objective lens inside the fused silica to generate modified micropatterns including the patterns of an embedded large-scale microchannel with 3D micro-mixing units and a string of vertical extra-access holes along the channel in fused silica. After laser modification, the sample was immersed in the chemical etching solution under the ultrasonic water bath to selectively remove all laser-modified materials. To achieve the homogenous fabrication of large-scale glass microchannels, the introduction of extra-access holes was adopted to break the inherent limitations of the selectivity of conventional ultrafast laser-assisted etching, enabling greatly enhanced etching performance of the microchannels[22-24]. Finally, the etched holes were sealed by defocused $CO_2$ laser irradiation on the glass surface to form the final large-scale fused silica microchannel structure. The defocused $CO_2$ laser irradiation provides a more spatially controllable sealing of those ports as compared with direct laser irradiation due to the improved intensity distribution for in-situ melting on the glass surface[22-24]. The spot size of the defocused $CO_2$ laser beam was larger than the dimension of the ports on the glass surface for rapid and controllable melting of the surrounding area of the holes. With optimized irradiation parameters, the holes could be sealed to form sealing layers with typical thicknesses of several hundred microns, which gives rise to the high-pressure resistance of the manufactured microchannel structures up to several MPa.

For high-performance UV photochemical synthesis, the substrate material of the microreactor is of vital importance for high-efficiency utilization of UV photon energies. To identify the superior advantages of fused silica for UV photochemical microreactors, transmittance curves of commercial borosilicate glass (the most commonly used materials for continuous-flow microchannel reactors) and fused silica with different thicknesses were measured and compared at multiple wavelengths ranging from 200 to 320 nm. As shown in the top panel of Figure 1b, for borosilicate glass samples with different thicknesses, there was strong absorption of UV photon energies. Moreover, with the increase in the thickness, the transmittance of the glass decreased severely. For instance, as presented in the bottom panel of Figure 1b, when the thickness of borosilicate glass increased from 1 to 3 mm, the corresponding transmittance value reduced from ~61.5% to ~14.4%. In contrast, the fused silica glass sample with different thicknesses exhibited high transmission rates at all the wavelengths in the range of 200~320 nm. Especially, as shown in the bottom panel of Figure 1b, the transmittance values of all fused silica samples were maintained to be above 90% at the wavelengths of 255, 275, 290, and 310 nm regardless of thickness. Therefore, to achieve a high photon utilization rate for UV photochemical synthesis, the adoption of fused silica as glass reactor material is highly preferable.

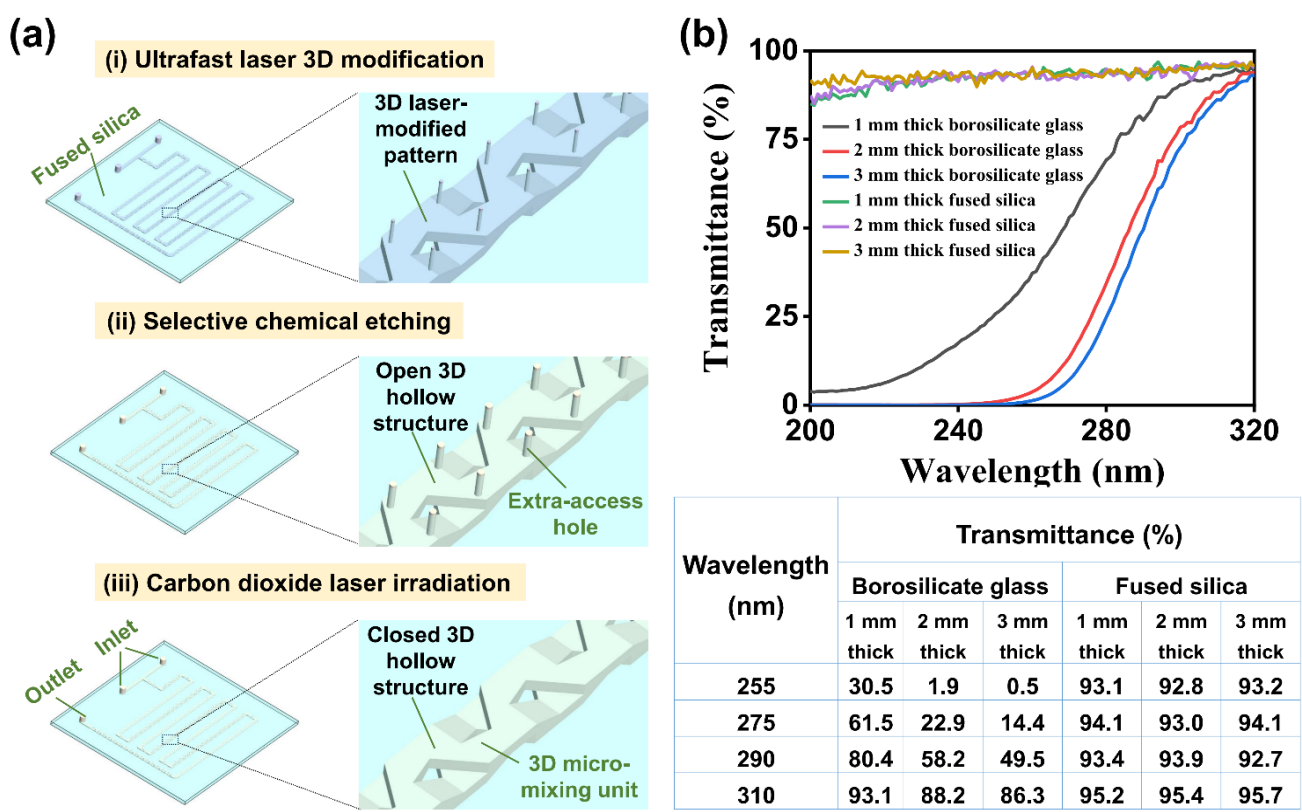

**Figure 1.** (a) Schematic of the manufacturing procedure for a 3D large-scale fused silica microchannel reactor, which consists of three main steps: (i) Ultrafast laser 3D modification; (ii) Selective chemical etching; (iii) Carbon dioxide laser irradiation. The right panel in each step presents a close-up view of the processed structure indicated by a dashed rectangle on the left panel. (b) Top: Transmittance curves of borosilicate glass and fused silica with different thicknesses at the wavelength from 200 to 320 nm;

Bottom: Comparison of the transmittances of borosilicate glass and fused silica with different thicknesses at the wavelengths of 255, 275, 290 and 310 nm.

## Construction of a continuous-flow UV photochemical synthesis system

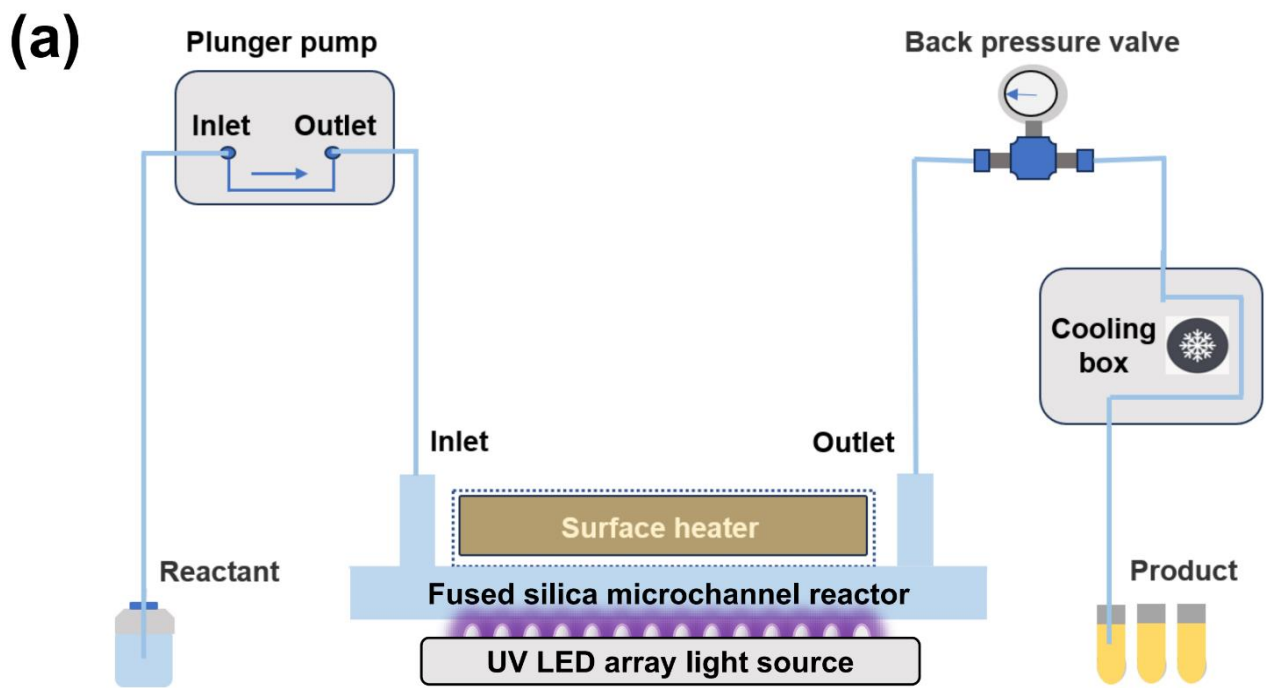

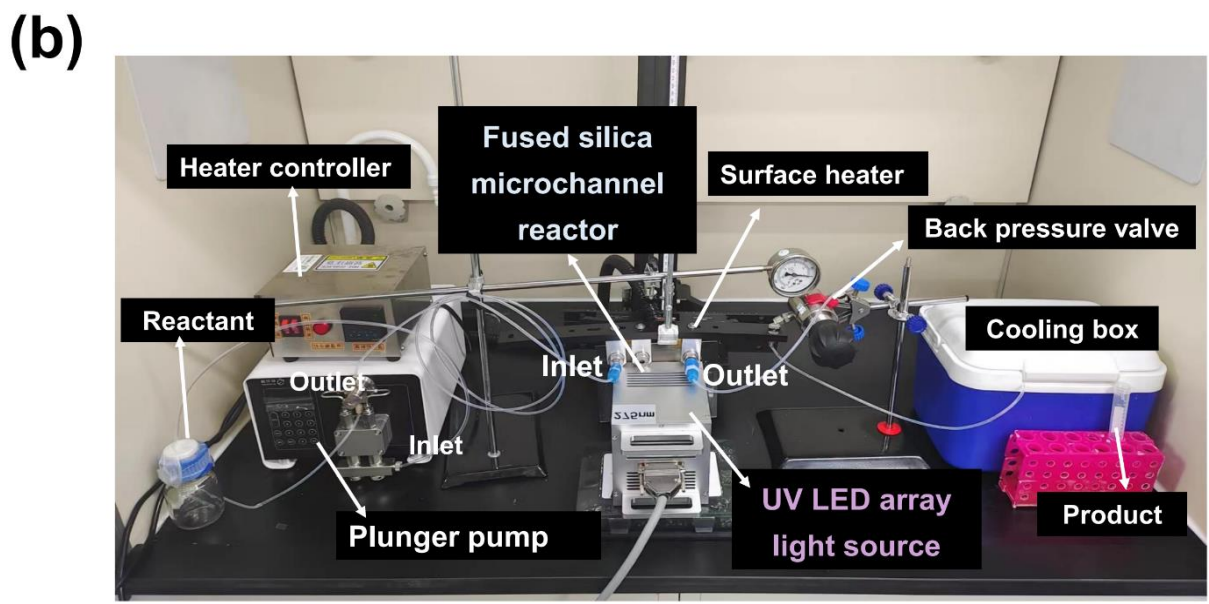

**Figure 2.** (a) Schematic and (b) photograph of continuous-flow UV photochemical synthesis system based on the fabricated fused silica microchannel reactor. The fused silica microchannel reactor was vertically placed between a surface heater and the UV LED array light source.

After the manufacture of the fused silica microchannel reactor, we built a continuous-flow UV photochemical synthesis system as illustrated in Figure 2a. This system comprises a plunger pump, a fused silica microchannel reactor, a UV LED array light source, a surface heater, an external back-pressure valve, and a cooling box. The inlet and outlet of the fused silica microchannel reactor were connected to the plunger pump and a back-pressure valve using polytetrafluoroethylene (PTFE) tubes, respectively. The bottom surface, i.e., the surface of the microreactor without the sealed access holes, of the microchannel reactor was attached to the surface heater while the top surface was attached to the UV LED light source. The entire device was of a sandwich layout in the order of LED-light source, fused silica microreactor, and surface heater. The interior temperature (available in the range of 25~350 °C) and pressure (available in the range of 0~3.5 MPa) of the microchannel reactor were controlled by the surface heater and back-pressure valve, respectively. The solutions of raw materials in a glass container were first pumped by the plunger pump into the microchannel reactor at the controlled back pressures and temperatures. Then, the on-chip photochemical reaction was performed under UV light irradiation. After flowing through the back-pressure value, the high-temperature reactants were cooled down to room temperature or even lower in the cooling box (the cooling medium could be liquid nitrogen, dry ice, ice water, etc.) before entering the final collection containers. The UV-LED array light source is composed of an array of UV-LED lamp beads at predesignated wavelengths. The irradiation power of the surface light source usually is determined by the power of a single lamp bead, the number of lamp beads per unit area, and the wiring density. Figure 1b presents a typical photo of a continuous-flow UV photochemical synthesis system in a fume hood. As compared with conventional UV photochemical systems based on batch-production methods, our home-built system provides superior performance for thermal management and UV light utilization of microreactor, which holds great potential for high-throughput and large-scale production.

**On-chip UV photochemical synthesis of Vitamin $D_3$**

In the UV photochemical microreaction system as shown in Figure 2b, the synthesis of VD3 includes the following procedures: First, the precursor solution was prepared by dissolving the

7-dehydrocholesterol (7-DHC) in methyl tert-butyl ether (t-BME)[5]. Secondly, the solution was pumped into the glass microchannel reactor with an interior pressure of ~2.0 MPa and a temperature of 160 °C. Once the pressure and temperature of the microchannel reactor were stabilized, the UV LED light irradiation was initiated for on-chip continuous-flow photochemical synthesis. Then, the reacted solutions were cooled down and collected for chemical analysis and characterization such as UV-vis absorption spectrum and high-performance liquid chromatography (HPLC). Herein, the influence of different continuous-flow photochemical synthesis conditions such as UV irradiation wavelengths and flow rates of reactant solutions on the synthesis performance of VD3 was systematically investigated.

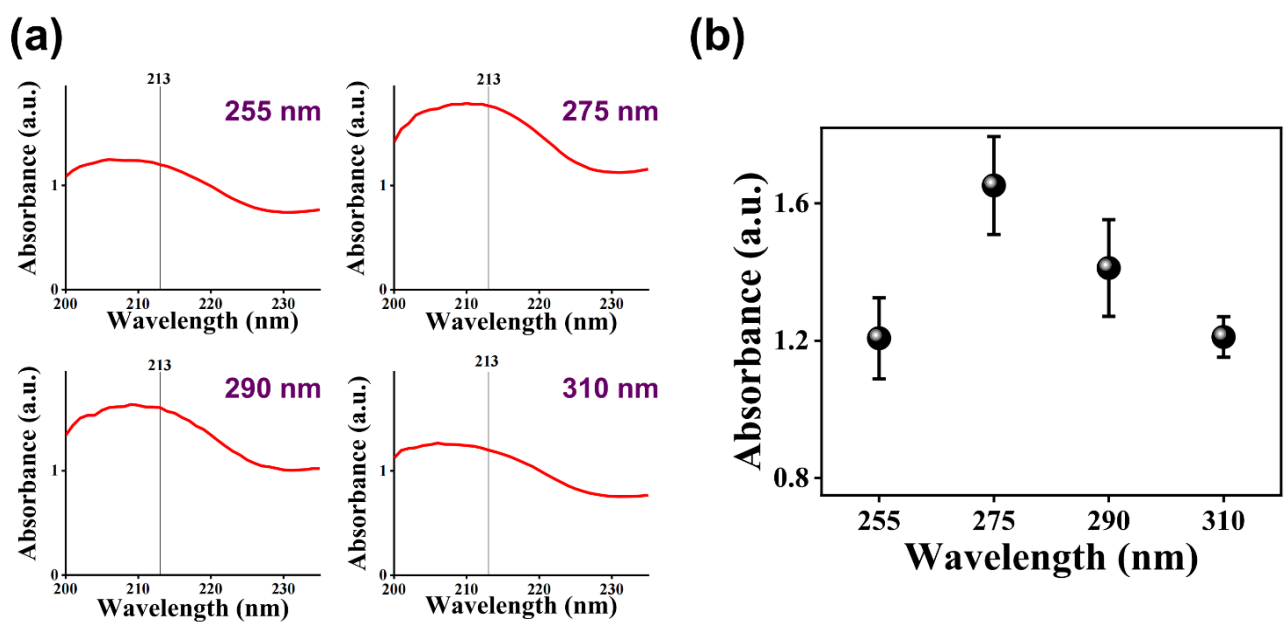

**Figure 3.** (a) Absorbance spectra of the synthesized products under the irradiation of UV LED array light sources at different UV irradiation wavelengths (255, 275, 290, and 310 nm) with a flow rate of 0.2 mL/min. (b) The intensity of the characteristic absorption peak at 213 nm of the products versus UV irradiation wavelengths in (a).

Figure 3a shows typical absorbance spectra of the synthesized products under the irradiation of UV LED array light sources (optical power: ~3.8 W) at different wavelengths (255, 275, 290, and 310 nm) with a flow rate of 0.2 mL/min (residence time: ~2.6 min). Two characteristic peaks in the absorption spectrum of a standard VD3 sample are observed at 213 and 264 nm (see Figure S1). These two characteristic peaks at 213 and 264 nm are used to determine the formation of VD3. However, it should be noted that the absorption at 264 nm is close to one of the characteristic absorption peaks of the standard DHC sample (see Figure S2). Therefore, using the intensity of the characteristic absorption peak at 213 nm for characterization is more precise. From the variation of the intensity of the characteristic absorption peak at 213 nm in Figure 3a, one can see that the corresponding UV irradiation wavelength with the maximum absorption intensity of the characteristic absorption peak of VD3 at 213 nm was 275 nm, which was of higher yield as compared with the UV irradiation sources at other wavelengths. To quantitively determine the influence of the irradiation wavelength on photochemical synthesis, the average values of the intensities of the characteristic absorption peaks at 213 nm were calculated and plotted in Figure 3b. Again, at the same irradiation powers of the LED array light sources, the 275 nm light source provides the maximum synthesized intensity among the different UV irradiation wavelengths (255, 275, 290, and 310 nm). To further confirm the successful synthesis of VD3, HPLC was employed to separate the synthesized products and detect the VD3 molecules. The emergence of the absorption at the retention time of ~6.9 min correlates to the formation of VD3 (see Figures S3 and S4). In contrast, the products obtained without UV irradiation show no strong absorption at the retention time of ~6.9 min, suggesting that the UV-induced photoreaction is responsible for the formation of VD3 (see Figures S4 and S5). Compared with conventional VD3 continuous-flow photochemical reactions and photothermal stepwise reactions[2-5], the one-step synthesis of VD3 reported in the current work had a yield of ~36% and a corresponding selectivity of ~41%. The high yield may be ascribed to the effective manipulation of reactant solutions in the UV-illuminated region in the fabricated microchannel reactor with 3D micromixing units[20, 24], high UV-photon-absorption efficiency of the reaction liquids in the reactor as well as the synergetic application of the high-pressure and high-temperature conditions[5].

    Figure 4a shows the absorption spectra of the products which were obtained under 275 nm LED array irradiation (optical power: ~4.2 W) at the different flow rates of reactant solutions (0.5, 1.0, 1.5, and 2 mL/min). The increase in the flow rate generally leads to a decrease in the residence time of the on-chip photochemical reaction, which tends to decrease the intensity of the characteristic peak at 213 nm as shown in Figure 4b. When the flow rate was raised to 2 mL/min, the average intensity of the absorption peaks of the products at 213

nm decreased to 0.8. Since the residence time is related to the total liquid inventory of the microchannel reactor, further improvement of the throughput and performance of UV photochemical synthesis can be achieved by increasing the volume size of the microreactor while maintaining the high mixing and heat transfer efficiencies, which will be the focus of our future investigation.

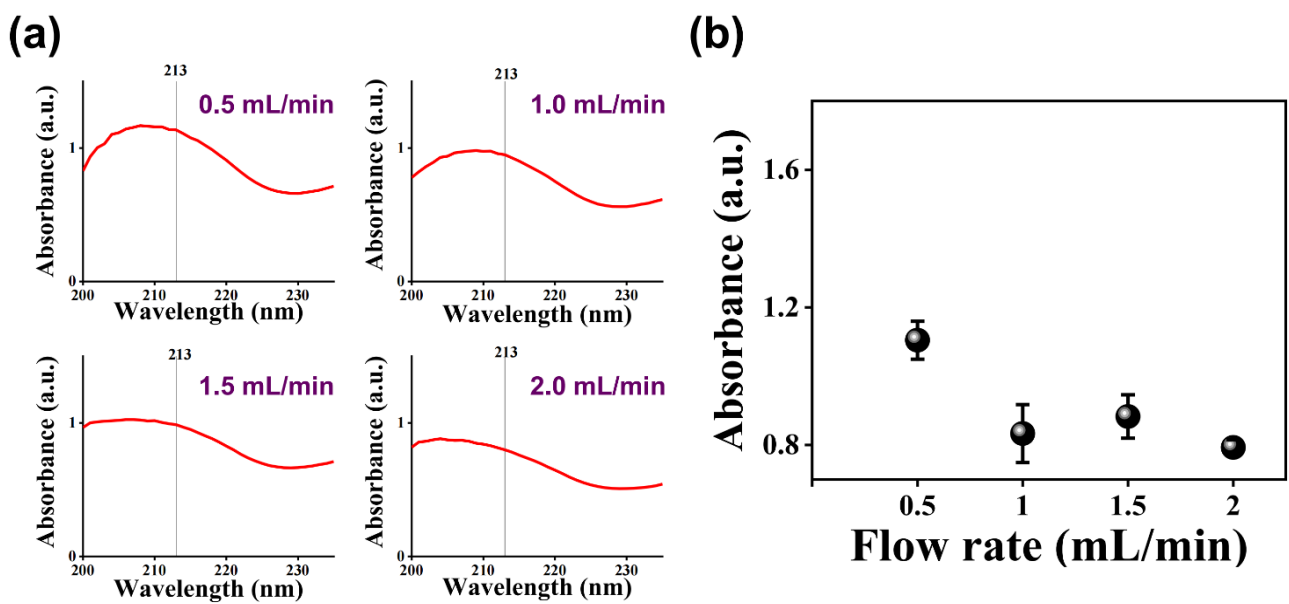

**Figure 4.** (a) Absorbance spectra of the synthesized products using different flow rates (0.5, 1.0, 1.5, and 2.0 mL/min) of reactants under the 275 nm LED array irradiation. (b) Intensity of the characteristic absorption peak at 213 nm of the products versus flow rate of the reactant solution in (a).

In summary, a one-step continuous-flow UV photochemical synthesis of VD3 is demonstrated in the 3D large-scale microchannel reactor fabricated using an ultrafast laser and a $CO_2$ laser. Compared with conventional UV photochemical microreactors, the proposed technique has the following advantages. First, the fabricated 3D fused silica microchannel reactor allows simultaneous operation of high-efficiency UV irradiation, high pressure, and high temperature for efficiency improvement in the continuous flow synthesis, which enables one-step and high-performance synthesis of VD3. The yield of synthesized VD3 reached ~36% and the corresponding selectivity was ~41%. Second, compared with traditional continuous-flow UV photochemical synthesis using high-pressure mercury lamps as radiation sources, the power consumption of photochemical synthesis based on UV LED array-assisted irradiation is greatly reduced[25,26]. Accordingly, the thermal effect of UV photochemical synthesis can be effectively controlled, which will be beneficial for further scale-up and mass production. Last but not least, compared with conventional visible photochemical synthesis heavily relying on expensive catalysts, the fused silica microreactor is expected to develop a series of catalyst-free UV photochemical synthesis strategies thanks to its unique capability of operating under high-energy UV photon irradiation.

## Materials and Methods

**Fabrication of fused silica microchannel reactors**

For 3D laser modification, an ultrafast laser amplifier system (Pharos 20 W, Light Conversion, Lithuania) with a central wavelength of 1030 nm, a tunable pulse width of 0.27~10 ps, a repetition rate of 250 kHz, and a maximum pulse energy of ~0.4 mJ was employed. The fused silica glass samples (JGS1) were amount on a 3D motorized stage (Coretech 3D X-Y-Z stage, Coretech, China) and irradiated by the focused laser beam through an infrared objective lens with a numerical aperture of 0.26 (M Plan Apo NIR 10X, Mitutoyo, Japan). The laser writing speed, line-by-line, and layer-by-layer writing spacing were 100 mm/s, 30 μm and 50 μm, respectively. For chemical etching, the laser-processed glass samples were immersed in a 10 M KOH etching solution at 95 °C under an ultrasonic bath until all modified regions were removed. For $CO_2$ laser irradiation, a defocused $CO_2$ laser beam (FSTi100SWC, Synrad, USA) was used to irradiate the extra-access ports one by one. The $CO_2$ laser beam was focused through a ZnSe lens (LA7028-E3, Thorlabs, USA) and then induced in-situ sealing on the openings of the extra-access ports on glass surface one by one with the translation of a 2D motorized stage (OneXY-500-500-AS-CMS1, Coretech, China).

**On-chip continuous-flow UV photochemical synthesis**

For on-chip photochemical synthesis, a precursor solution with a concentration of 21 mg/mL was prepared by dissolving 7DHC in t-TBE. The 7DHC solution was driven by a plunger pump (DPS-100, Oushisheng, China) into the inlet of the chip, and the chip with a size of 155 mm × 125 mm × 2 mm was attached to the UV LED light source and the heater, forming a triple-layer sandwich structure so that the UV light irradiation and the reaction temperature of the channel reactor could be simultaneously controlled for photo- and thermo-reactions. The illumination areas of all UV LED light sources were 100 mm × 100 mm. The volume of the fabricated microchannel reactor in the UV-illuminated areas was estimated to be ~0.53 mL. During the on-chip synthesis, the temperature of ~160 °C and the pressure of ~2 MPa in the microchannel reactor were kept constant. The collected products were characterized by a UV absorption spectrometer (UV-2600, Shimadzu, Japan) and a high-performance liquid chromatography (Agilent 1100, Agilent Technologies, USA).

## Acknowledgments

This work was supported by the National Natural Science Foundation of China (Grant Nos. 12174107, 61991444, 11933005, 12192251, and 11734009); National Key R&D Program of China (Grant No. 2019YFA0705000); Science and Technology Commission of Shanghai Municipality (Grant No. 21DZ1101500); Shanghai Municipal Science and Technology Major Project.

## Author contributions

Y. Cheng supervised the research work. J. Xu, M. Hu, and Y. Cheng conceived the experiments. A. Zhang, J. Xu, L. Xia, M. Hu, Y. Song, and M. Wu carried out the experiments. A. Zhang, J. Xu, and M. Hu contributed to the data analysis. All authors participated and contributed to the writing of the manuscript.

## Conflict of interest

The authors declare no competing interests.

## Supplementary information

Supplementary materials are available in the online version.